**Title:** Impact of Synthetic Lesional MR Images in Automated Focal Cortical Dysplasia Detection in Low-Data Scenarios

**Running title:** Synthetic MRI for FCD Detection

**Authors:** Prabhjot Kaur; Hakim Ouaalam; Sedat Kandemirli; Sanjay P. Prabhu; Simon K. Warfield

**Institutional Affiliations for all authors:**

Computational Radiology Laboratory, Boston Children's Hospital, Harvard Medical School, Boston, MA, USA

**Corresponding Author:**

Prabhjot Kaur

Computational Radiology Laboratory

Boston Children's Hospital, Harvard Medical School

300 Longwood Avenue, Boston, MA 02115, USA

Tel: +1 (857) 318-6057

Email: prabhjot.kaur@childrens.harvard.edu

**Keywords**: focal cortical dysplasia; synthetic MRI; epilepsy; lesion detection; generative AI

**Funding**: This research was supported in part by NIH S10 OD025111, R01 LM013608, R01 EB019483, R01 NS124212, and by an award from the Thrasher Research Fund.

**Abstract:**

**Background and Purpose:** Automated detection of focal cortical dysplasia (FCD) requires large volumes of voxelwise-lesion delineated MRI data, which are difficult to acquire. This study aims to generate synthetic MRI data exhibiting FCD, assess their realism, and evaluate their impact on automated FCD detection—particularly in reducing the need for manual annotations.

**Methods:** T1-weighted (T1w) and T2-weighted-Fluid-Attenuated Inversion Recovery (FLAIR) MRI scans from 131 FCD patients and 90 healthy controls from multiple (3) sites, were retrospectively studied. Synthetic MRIs were generated by conditioning a generative network on binary FCD mask. Two neuroradiologists identified real images from a random set of 14 real and 14 synthetic scans. Three nnU-Net models were trained to detect FCD using: (i) real-only (35-FCD/35-controls), (ii) real (35-FCD/35-controls) + synthetic augmentation, and (iii) expanded real data (70-FCD/70-controls).

**Results:** Experts showed limited ability to distinguish real from synthetic images, with classification accuracy of 60% for T1w and 70% for FLAIR (inter-rater agreement $\kappa = 0.86$). Augmenting automated FCD detection with synthetic data increased sensitivity by 8.14% ($p = 0.12$) and improved model confidence at true lesion sites ($0.83 \pm 0.11$ to $0.89 \pm 0.12$; $p = 0.02$). The expanded real-data model further improved sensitivity to 73.8% ($p < 0.001$) and confidence to $0.90 \pm 0.14$ ($p = 0.01$).

**Conclusion:** Conditional generative networks can generate realistic synthetic FCD-MRIs, reducing labeled data needs by ~20% while maintaining equivalent sensitivity. Equivalent amounts of real data, when available, remain more effective than synthetic augmentation.

## INTRODUCTION

Focal Cortical Dysplasia (FCD), a common cortical malformation often overlooked during manual MRI review, is the leading cause of drug-resistant epilepsy (DRE) in children and the third most common cause in adults [1–3]. Early detection of cortical malformations causing drug-resistant epilepsy (DRE) is critical for timely surgical treatment and better outcomes [1]. Typically, expert radiologists visually inspect MRI scans to identify abnormal regions. When detected on MRI, FCD is associated with seizure freedom rates of up to 70%, compared to ~40% in MRI-negative cases [4]. However, nearly 30% of Focal Cortical Dysplasia (FCD) cases are missed by radiologists due to subtle features, perceptual errors, or poor MRI scan quality [4–5].

Machine learning (ML) and deep learning (DL) approaches have recently been proposed to assist in manual FCD detection from MRI, acting as secondary readers to support radiologists and reduce perceptual errors [6–10,14]. These models typically rely on supervised learning using MRI images paired with delineation of lesions for each voxel. The lesions were manually delineated voxel-by-voxel by expert radiologists and neurosurgeons to indicate abnormal regions exhibiting FCD, which are referred to as FCD labels in this work. FCDs exhibit the following MRI imaging characteristics: gray-white matter junction blurring, cortical thickening, transmantle sign, abnormal gyration, and increased signal intensity on T2-weighted or FLAIR sequences [4]. However, manual voxel-wise annotation is time-consuming and labor-intensive, limiting the availability of large annotated datasets. The collection of large dataset demands for multi-site data sharing which introduces logistical and regulatory barriers, often requiring significant time and effort [4]. Consequently, the development of current methods is constrained by both the scarcity and limited accessibility of labeled data [4].

Generative deep learning is an approach that improves the performance of tasks such as image segmentation and anomaly detection - where annotated datasets are limited [19]. This approach estimates the underlying distribution of training data to synthesize new and realistic data samples. These synthetic data are then combined with limited annotated datasets to expand both the size and diversity of training data, without requiring additional manual annotation [11]. Recent studies have demonstrated the use of generative models to create annotated pathological brain images, effectively augmenting small datasets in applications such as brain tumor and white matter hyperintensity detection [12–13,17].

However, this generative approach has not yet been explored for producing synthetic annotated images exhibiting FCD or evaluating its impact on FCD detection in data-scarce scenarios. The direct application of generative modeling in FCD detection is challenged by the limited availability of labeled FCD cases. For instance, approximately 800 annotated MRI scans were required in prior generative based studies detecting brain tumors or white matter hyperintensities [13]. This is comparable to the largest curated multi-site dataset for FCD detection to date, comprising around 762 subjects [9]—a significant effort that highlights how such datasets are not readily available or accessible. The need for large annotated datasets in [12–13, 33-34] arises from the use of highly parameterized latent diffusion models and variational encoder-decoder networks, trained to generate both tissue probability maps and binary FCD labels.

Furthermore, current methodologies align with group-theoretic frameworks where augmentation acts as a variance-reduction tool over invariant data orbits [31]. While standard augmentation techniques—such as rotation, scaling, or flipping— ensure the model is invariant to orientation, they cannot simulate the complex, spatially-adaptive intensity changes characteristic of FCD lesions [33]. In contrast, domain randomization strategies utilize anatomical maps to ensure model robustness against cross-site scanner variability [32]. Consequently, there remains a critical methodological gap for a framework that can synthesize pathologically diverse data without the massive sample sizes required by unconstrained diffusion models.

To address the above challenge, we propose embedding FCD like regions in healthy brain scans providing MRI images exhibiting FCD along with corresponding binary FCD labels. This strategy eliminates the need to generate brain scan from scratch - previously done using diffusion models in [12-13]. The elimination of diffusion models reduces trainable-parameters (by 95%), and thus lowers dependence on large annotated datasets. To our knowledge, this is the first study to introduce a generative framework for FCD detection in low-data scenarios, addressing a critical gap in the literature. The objectives of this work are to: (i) generate synthetic MRI volumes exhibiting FCD, (ii) augment it with training data, and (iii) evaluate the impact of this augmentation on automated FCD detection. Our key contributions include: (i) a novel GMM-based method to generate anatomically plausible FCD labels from healthy brain scans, (ii) application of a conditional generative model to synthesize MRI with FCD, and (iii) evaluate impact of this augmentation on automated FCD detection. We hypothesize that enriching training data with synthetic FCD-labeled volumes will improve detection accuracy.

## METHODS

### 1. Dataset:

We utilized three imaging datasets: two publicly available and one institutional. The number of patients/controls, histopathology, scanner information, acquisition and spatial resolution, present in datasets are described in Table 1.

Public datasets: To assess the voxel-wise delineation of FCD, a dataset with MRI T1w and T2w-FLAIR volumes along with voxel-wise manual delineation of FCD is required. The large dataset collected for epilepsy with FCD in MELD [7] is publicly available only with surface based features and hence restricts the volumetric analysis required in this work. In contrast, the only open-source volumetric FCD dataset currently available is the Bonn dataset (n=85), which is roughly ten times smaller than typical tumor datasets. We used Bonn dataset - dataset-1 (D1) [16] and Dataset-2 (D2) [15] that are publicly available dataset and were accessed in June 2024 under a CC0 license. Subjects with histopathologically proven FCD type-II regions were selected from both datasets. The public datasets provided manual lesion annotations by expert neuro-radiologists, which guided surgical resections, and are considered as the ground truth in this work.

Institutional dataset: Dataset-3 (D3) was obtained from the epilepsy program at Boston Children's Hospital (BCH), a tertiary referral center specializing in pediatric epilepsy surgery. The study was approved by the Institutional Review Board in July, 2020 with waiver of informed consent, and the data was handled in compliance with Health Insurance Portability and Accountability Act regulations.

*Inclusion criteria:* Subjects who underwent surgery between 2015 and 2023 with histopathologically confirmed FCD type-II and who achieved at least two years of postoperative seizure freedom.

*Exclusion criteria:* Subjects with incomplete records, less than two years of follow-up, or ambiguous clinical or histological findings.

This constitutes a larger set, and we randomly sampled a subset of 14 patients for this work. We also randomly chose 5 control subjects with no structural abnormality to evaluate the specificity of methods. Patients with Engel class-I outcomes were included, and manual lesion segmentation was performed using post-surgical resection cavities as guidance, ensuring that labeled regions corresponded to the presumed focal cortical dysplasia (FCD).

All MRI scans were skull-stripped using FreeSurfer [21] to limit analysis to brain tissue and prevent non-brain structures from affecting intensity normalization [24]. Partial volume estimates (PVE) represent the fraction of a tissue present in the given voxel. PVE for six brain tissues: GM, WM, CSF, deep GM, brainstem and Cerebellum - required for generating synthetic lesional volumes, were obtained from brain masks generated with Geodesic information flows and niftyseg libraries [20].

2. Proposed Method:

(i) FCD Label Generator with GMM: The initial step in our methodology involves the FCD Label Generator, which produces a voxel-wise binary mask delineating the precise presence and spatial extent of the FCD lesion. This generator is informed by the characteristic imaging appearance of FCD Type II lesions: blurring of the gray matter-white matter (GM-WM) junction and a typically smooth spatial extent.

For a given healthy MRI image with $I(x)$: MRI intensity at voxel $x \in \mathbb{R}^3$, the PVE maps for GM-WM as ($PVE_{GM}$ and $PVE_{WM}$) were computed such that $PVE_{GM} + PVE_{WM} + PVE_{CSF} = 1$. The set of boundary voxels were calculated as

$$\mathfrak{B} = \{ x \in \mathbb{R}^3 \mid |PVE_{GM}(x) - PVE_{WM}(x)| < \varepsilon, PVE_{GM} > 0.3, PVE_{WM} > 0.3, PVE_{CSF}(x) < 0.05\}$$

Here, a small threshold ε (set to 0.1 in this work) was introduced to ensure that $\mathfrak{B}$ contains pixels with approximately similar fraction of WM and GM. We randomly selected a voxel $\mu_0 \in \mathfrak{B}$ as the center for lesion generation, $\mu_0 \sim \text{Uniform}(\mathfrak{B})$. This voxel served as the reference mean for constructing a Gaussian Mixture Model (GMM).

The number of mixture components (K) was chosen at random, $K \sim \text{Uniform}\{1, 2, 3, 4, 5\}$. For each component, the mean offset $\mu_k$ was sampled independently from a discrete uniform distribution over [1,5], and the covariance matrices $\Sigma_k$ were drawn from a random positive-definite distribution to produce varying shapes and orientations. The GMM probability density at voxel x was defined as: $p(x) = \sum^{K}_{k=1} \pi_k \cdot \mathcal{N}(x \mid \mu_k, \Sigma_k)$, where π□ denotes the mixture weights with $\sum^{K}_{k=1} \pi_k = 1$. This formulation allows the GMM to generate smooth yet irregular and anatomically plausible lesion boundaries.

The probabilistic lesion map was then multiplied by a combined GM+WM mask (i.e., a union of GM and WM labels) to constrain lesion generation to tissue types relevant to FCD. Randomizing lesion location, extent, and morphology provides a diverse set of synthetic FCD labels across the brain.

(ii) Lesional volume generator with SPADE-GAN:

To generate T1-weighted (T1w) and T2-weighted FLAIR (T2w-FLAIR) images from brain tissue labels and FCD labels, we utilize a semantic image synthesis framework conditioned on label information. Traditional conditional GANs often concatenate label maps with the input at the initial layer and treat them as intensity images. In convolutional neural networks (CNNs), however, this semantic information is typically downsampled and normalized, leading to loss of spatial precision and semantic structure degradation.

To address this, we adopt the SPatially-Adaptive DEnormalization (SPADE (v1.0, NVIDIA Corporation, Santa Clara, CA, USA, https://github.com/NVlabs/SPADE) based Generator with variational autoencoder (VAE) architecture embedded with discriminator [12], which preserves semantic details at multiple layers of the VAE. At a given layer l, the feature activation $f_l(x) \in \mathbb{R}^{C \times H \times W}$ , was first normalized using BatchNorm: $\hat{f}_l(x) = (f_l(x) - \mu_l) / \sigma_l$ where $\mu_l$ and $\sigma_l$ are the mean and standard deviation of the features at layer l, and C, H, and W denote channels, height, and width, respectively.

Then, the normalized features were modulated using affine parameters γ_l(s), β_l(s) derived from the semantic label map s (which includes FCD label and PVE maps):

$$f_{SPADE_l}(x, s) = \gamma_l(s) \cdot \hat{f}_l(x) + \beta_l(s)$$

Here, $s \in \{0,1\}^{K \times H \times W}$ is a multi-channel binary semantic map representing K channels, K=7 brain tissue classes and FCD regions. The affine parameters were generated through a small CNN applied to the resized semantic label map at each layer l, thus preserving semantic consistency at every resolution.

In addition to semantic guidance, we introduced style conditioning by sampling a style image $I_{style}$ representing the desired MRI modality (T1w or T2w-FLAIR). This image was encoded using a separate style encoder E to produce a latent code $z = E(I_{style})$, which modulated the overall appearance of the output image.

The final generator output is given by: $\hat{I}_{gen} = G(s, z)$ where G is the SPADE-based generator conditioned on the semantic label map s and style code z. The loss function includes an adversarial loss $L_{GAN}$, a perceptual loss $L_{VGG}$, and a style consistency loss $L_{style}$ to enforce visual and modality fidelity.

An example of this semantic-to-MRI translation, where a healthy MRI image and FCD label are combined to synthesize an FCD-expressing image, is shown in Figure 1.

(iii) FCD Detector with nnUnet:

The DL model with nnU-Net architecture is a standard baseline for segmentation tasks given its minimal need for manual tuning of data preprocessing steps and network hyperparameters [18,13]. Its reliability across datasets have been well documented in the literature [22]. As part of this study, we systematically compared the nnU-Net–based approach with the MELD project's method, trained on the largest known FCD dataset [7]. Despite using a smaller training set, the nnU-Net-based-method achieved a superior performance than MELD [7], see Table 4 (Methods 1-3).. It led us to adopt the nnU-Net–based-method as our preferred method for automated-FCD-detection.

The nnU-Net model was trained using the default parameters described in [18], employing a composite loss function that combined Dice loss and cross-entropy (CE) loss at each decoder level, as in [18]. This formulation balances region-overlap performance with voxel-wise classification accuracy.

(iv) Model Training:

Lesional volume generator: The model was trained to provide T1w or T2w-FLAIR image depending on the style image. This model was trained for 200 epochs, and the shape of input was [7x128x128x64] where 7 channels include PVE for 6 brain tissues, and 1 FCD binary label. The MR volumes were cropped to 128x128x64 patch size for training with an overlap of 10 voxels. The architecture of the SPADE model is detailed in [https://github.com/virginiafdez/brainSPADE3D_rel].

FCD Detector: Trained from scratch, the network utilizes co-registered T1w and T2w-FLAIR images as a two-channel input. The training set was constructed by cropping the input volumes and FCD label into 128x128x64 cubes with a 10-voxel overlap in each direction. The model predicts FCD locations, which are compared directly against the paired manual labels. We employed the '3d_highres' configuration to avoid loss of image-details and batch size of 2. The model was trained for 300 epochs (negligible improvement in training dico score after 300 epochs) for 5 different folds of training and validation. The learning rate was 0.01.

3. Experiment Design:

The objective of this study is to investigate both the generation of synthetic FCD-containing MRI volumes and assess the impact of synthetic data on automated FCD detection performance. To support this objective, we designed following evaluations:

(i) Subjective Quality Assessment: We conducted a blind classification experiment where two expert raters evaluated the realism of synthetic data. One of the experts is a fellowship trained board certified pediatric neuro-radiologist with experience of 21 years and another expert is fellowship trained board certified pediatric neuro-radiologist with post-residency experience of around 5 years. Experts classified 28 total images across two categories: T1w and T2w-FLAIR. Each category contained 14 images (7 real and 7 synthetic) that were independently and randomly sampled. For every image, the experts had to determine if the scan was real or synthetic. The discrimination is evaluated using accuracy and inter-rater agreement.

(ii) Impact of synthetic data on FCD Detection Performance: We evaluated the effect of synthetic data augmentation by comparing three FCD detection models: FCDD-1, trained on a limited real dataset without augmentation; FCDD-2, trained on the same real data augmented with synthetic volumes; and FCDD-3, trained on an expanded real dataset matching FCDD-2 in size, serving as an upper-bound reference—illustrated in Figure 2.

*(ii. A) Construction of training and Test Cohorts:* Three FCD detectors (FCDD) were trained on different datasets: FCDD-1 used 35 real FCD and 35 healthy images; FCDD-2 used the same real data augmented with 35 synthetic FCD and 35 synthetic healthy images; and FCDD-3 used an expanded real dataset of 70 FCD and 70 healthy subjects, matching FCDD-2 in size. The synthetic data for FCDD-2 were generated using SPADE trained on the same 35 FCD and 35 healthy volumes used in FCDD-1. The reduced real-data cohort size of 35 FCD and 35 healthy subjects was selected based on supplementary analyses (Supplementary Table 1A and Figure 1A) across multiple training-data proportions, which showed that this was the minimum amount of real data at which synthetic augmentation became beneficial in FCD detection.

*Test Set 1 (Held-out from Dataset-1):* The remaining 15 unseen FCD and 15 healthy cases were used for testing. These were selected to ensure variation in lesion size (1,125–15,177 $mm^3$), location (frontal, parietal, temporal), hemisphere (left/right), and FCD subtype—8 Type-IIb, 3 Type-IIa, and 4 with unspecified subtypes.

*Test Set 2 (External Sites):* To assess generalizability, we included two other datasets (D2 and D3) collected at different institutes for evaluating the method, comprising 46 patients: 32 from D2 and 14 from D3; 5 controls from D3. The 15 patients and 15 controls from D1 were also included in this test set 2 that were also in test set 1.

The subjects from Test set 1 and Test set 2, were not used in any manner for training any of the methods in this paper.

(ii.B) Performance Metrics: A predicted FCD location was considered a true positive if at least one voxel overlapped with the expert-labeled FCD region, and is reported in detection rate. In addition to the predictions that overlap with expert-labeled masks, the predicted clusters located within 26$mm^3$ neighbouring lesional voxels in ground truth but without direct overlap were also included in the pin-pointing rate metric for subject-level analysis. False positives were defined as regions predicted as abnormal by the method but labeled as normal in the ground truth. Additional performance metrics—accuracy, sensitivity, positive predictive value (PPV), negative predictive value (NPV), F1-Score, and specificity—were used to compare different models.

(iii) Comparison with existing methods: We selected nnU-Net for lesion segmentation in MRI based on prior comparisons with existing FCD detection methods [6–7,9,13]. We contrasted these approaches in terms of training data requirements, number of trainable parameters, test set size, accuracy, sensitivity, and specificity. Additionally, we evaluated the effect of synthetic data augmentation with limited training data against the performance achieved using the largest available public dataset on the D1 test cohort.

(iv) Statistical analysis

Data analysis was performed using SPSS 18.0 (IBM, New York, USA) [23]. Continuous variables are reported as means with standard deviations (SD) or as medians with interquartile ranges (IQR) or 95% confidence intervals (CI). Categorical variables are summarized as frequencies and percentages.

## RESULTS

(i) Subjective evaluation of synthetically generated lesional volumes:

Figure 3 illustrates the quality of the synthesized images, while Figure 6 maps the spatial distribution of the generated images. The observed divergence—particularly the increased representation of less frequent lesion locations—is a consequence of our framework not imposing any a priori assumptions on lesion distribution, other than requiring at least one voxel to overlap with the gray matter–white matter boundary. This design was intended to reduce representation bias toward lesion lesions prevalent in the training data, thereby promoting diversity of the augmented training set with underrepresented lesion presentations. In addition, lesion sizes were constrained (maximum ~5k $mm^3$) to promote oversampling of smaller lesions, which are clinically more challenging to detect. An average accuracy of 60% for T1-weighted (T1w) scans and 70% for FLAIR scans was achieved. Interrater agreement was high (Cohen's

kappa = 0.86; 95% CI: 0.69-1.00), indicating that the synthetic images were highly realistic and often indistinguishable from real scans.

(ii) Impact of synthetic lesional volumes on FCD Detection:

*Test Set 1 (Single-site evaluation):* When evaluated on Test Cohort 1 from the same site as the training data, synthetic data augmentation improved sensitivity and reduced the false positive rate compared with FCDD-1 (trained without synthetic data). The performance metrics are detailed in Table 2.

*Test Set 2 (Multi-site evaluation):* With synthetic data augmentation, five additional subjects were correctly identified on the subject level compared to the baseline model without augmentation. The performance metrics for each of FCDD1, FCDD2 and FCDD3 are detailed in Table 3. The subject-level sensitivity for FCD detection in the full test cohort (n = 61) as mentioned in Tables 2 and 3, was increased modestly from 36.1% (22/61) without augmentation to 44.3% (27/61) with augmentation—an absolute improvement of 8.2%, p=0.22 (95% CI: 25.2%–48.6% vs. 32.5%–56.7%, $p$ = 0.12). Additionally, the model's predicted probabilities at true lesion sites improved significantly with synthetic data, rising from 0.83 ± 0.11 to 0.89 ± 0.12 ($p$ = 0.02), indicating increased model confidence at lesion locations. This performance was equivalent in terms of sensitivity and specificity achieved by nnUNet trained on 20% more real data (42 controls/ 42 FCD).

FCDD1 showed a bias toward classifying volumes as normal, resulting in nine missed FCD cases. Of these, five were correctly identified by FCDD2; however, this improvement was accompanied by four false positives. Example cases where FCDD1 and FCDD2 produced conflicting predictions are illustrated in Figure 4.

In comparison, training the model with an equally sized expanded set of real MRI data yielded a significantly higher detection sensitivity of 73.8% (45/61; 95% CI: 61.6%–83.2%; $p < 0.001$) and further increased model confidence (0.90 ± 0.14; $p = 0.01$). These results demonstrate that while synthetic data can improve model performance, equal sized real data remains more effective when available in similar amounts.

(iii) Comparison with existing methods: As compared to MELD [7], the proposed method with augmentation achieved higher accuracy on Dataset-1 (0.73 vs. 0.40) while using 84% less real data, see Table 4. The false positive rate decreased from 1.3 to 0.7 in the proposed method as compared to MELD [7], example shown in Figure 5. Sensitivity improved by 13.3 percentage points in Dataset-1 (73.3% vs 60.0%) for proposed method with augmentation and 7.1 percentage points in Dataset-3 (64.3% vs 57.1%) for proposed method real-expanded approach, neither reaching statistical significance (McNemar's test, $p=0.50$ and $p=1.00$, respectively). Performance metrics are detailed in Table 4.

The proposed generation method, leveraging GMM and healthy cohorts, reduced the number of trainable parameters (the weights and biases learned during model training) by 95%, thereby lowering labeled data requirements compared to the baseline method [13] used for developing the conditional generation method in this study.

## DISCUSSION

Accurate detection of Focal Cortical Dysplasia (FCD) remains challenging, particularly in data-scarce settings and suboptimal imaging. This study shows that synthetic data augmentation can improve automated FCD detection, even with limited annotations. The synthetic images closely resembled real MRIs and enabled detection of additional FCD cases previously missed. While sensitivity gains were modest, lesion-level confidence improved significantly. However, augmentation also increased false positives—nine cases were missed without augmentation, while four false detections occurred with augmentation—highlighting both its potential and need for refinement. Compared to existing methods, our approach performs competitively with fewer training samples.

Increasing the number of training samples does not always improve FCD detection performance, as observed in MELD [7], and recent studies have reported similar performance trends for MELD [25-26]. Previous work [22,27] also showed improved detection when switching to nnU-Net without increasing data. In this study, we showed that for a fixed nnU-Net model, performance decreased with reduced real labeled data. Additionally, our method demonstrated the feasibility of using synthetic data to reduce training data requirements while improving performance in low-data settings. This opens a new research direction that leverages emerging technologies to further improve FCD detection.

Our proposed framework leverages publicly available healthy datasets, which inherently eliminates the need for collecting additional patient data or manual lesion annotations. This approach offers a significant advantage over other data generation strategies. For instance, methods that utilize semantic segmentation conditioned on synthetic lesions, such as diffusion models or GAN used to generate tumor images, typically require large training sets (e.g., 850-1500 subjects) and often rely on unstable techniques like shifting and flipping real tumor labels [13, 29]. Similarly, frameworks like SynthSeg [30] have demonstrated that training on purely synthetic images—generated from label maps using Gaussian Mixture Models (GMM)—can achieve high anatomical segmentation accuracy across diverse image contrasts. An alternative trend involving using a segmentation network to generate pseudo-labels which then train a separate generation network in ultra low data regimes[28], remains a valuable area for future research. Importantly, the proposed work provides a controllable way to provide diversity on lesion location and lesion size (as shown in Figure 6) by controlling the hyperparameters of GMM unlike diffusion models based methods [13].

This work demonstrated a clear performance hierarchy: real data outperforms an equivalent volume of synthetic data, yet the inclusion of synthetic data significantly improves performance over a non-augmented baseline in low resource settings. A crucial area for future work is determining the upper bound of improvement achievable by substantially increasing the synthetic data volume (e.g., up to 100 times the real data, consistent with trends in computer vision). Unlike the current work, where validation was limited to embedding FCD only in small number of real healthy scans, the future investigation will utilize large public healthy datasets.

The current study utilized two distinct evaluation frameworks, Test Set 1 and Test Set 2, to determine if GAN-based data augmentation offers performance benefits beyond merely mitigating subject-level overfitting. Specifically, Test Set 1 assessed model robustness across individual subject variability, while Test Set 2 evaluated the framework’s resilience to multi-center scanner effects. Our observations indicate that augmenting real data with synthetic samples has a minimal effect on detection performance once the real-world data reaches a sufficient scale. In our setup—comprising 70 healthy controls and 70 patients—the model captured enough biological variance that additional synthetic data provided no measurable benefit for the unseen test cohorts. This suggests that while synthetic augmentation is critical for generalizing across disparate imaging sites in low-data regimes, its utility diminishes as the real-world training dataset achieves sufficient scale.

This study has several limitations. First, the quality of generated images depends on the accuracy of PVE maps, with CSF estimation often unreliable in low-quality scans, causing blurring. Second, only a limited number of synthetic examples—a subset of 35 healthy subjects from the same dataset (Dataset-1)—was used to generate synthetic lesional images, matching the 35 additional real lesional cases used to train FCDD-3. While this controlled design supported fair evaluation, it limited exploration of larger-scale synthesis and multi-site acquisition diversity. In future work, we aim to increase the number of synthetic volumes, and improve the quality of synthetic images by incorporating prior knowledge of healthy brain regions to guide lesion-focused generation, thereby improving site-level generalization, preserving tissue boundaries, reducing blur, thereby, further decreasing requirement of labeled real data.

This approach is currently limited to FCD type-II due to their accessibility in public datasets and typically learns to generate FLAIR hyperintensity as largely present in FCD Type-II training data. In future, it can also be applied to TSC subjects and other anomalies, including FCD type-I, which are rare and therefore particularly well suited for generative methods. Both proposed methods demonstrated substantial accuracy improvements over MELD-1 across datasets. The 'real-expanded' method demonstrated improved sensitivity over MELD on both datasets (+0.20 on Dataset-1, +0.07 on Dataset-3), while the 'synth+reduced real' method showed +0.13 improvement on Dataset-1 but a -0.07 decrease on Dataset-3. The reduced performance on Dataset-3 for the synth+reduced real method may reflect limited cross-site generalization due to training on only 35 subjects compared to MELD's ~200+ subjects. Despite this, both proposed methods achieved substantially higher accuracy than MELD on both datasets. Post-hoc power analysis indicates approximately 164-412 and 1,530-1,592 patients are needed to achieve statistical significance ($\alpha$=0.05, power=0.80) for the observed sensitivity differences in Dataset-1 and Dataset-3, respectively. Another assessment that is currently limited by the small number of informative cases includes: augmentation improved detection in two FCD Type-IIa cases (parietal ~2k $mm^3$ and frontal ~5k $mm^3$), whereas one FCD Type-IIb case (~5.2k $mm^3$, frontal) was detected without augmentation but missed with augmentation in dataset-1. This analysis will also require a larger patient cohort to determine whether augmentation benefit is influenced by the location, size, or imaging appearance of synthetic lesions.

**Acknowledgements and Disclosure:** A part of this work was presented as an oral presentation at the 2025 Annual Meeting of the American Society of Pediatric Neuroradiology (ASpNR), where it received the Best Oral Presentation Award. This research was supported in part by NIH S10 OD025111, R01 LM013608, R01 EB019483, R01 NS124212, and by an award from the Thrasher Research Fund.

**Table 1.** Summary of subject count, demographic information, and acquisition details for the three datasets.

| Data set | Scanner Strength and Model | Patients/ Controls | FCD Type distribution | Number of epilepsy patients | Modality | Age (min-max) | TR / TE / TI [ms] | Resolution (x,y,z) [mm] |
|---|---|---|---|---|---|---|---|---|
| D1 | 3T, Siemens Prisma | FCD Patients | Type IIa: 32%, Type IIb: 68% | 85 | T2w-FLAIR | 12 to 65 yrs, 28.9 ± 12.4 | [5000-7000] / 388 / [1800-2100] | 91.8% data is 3D FLAIR: 1x1x1, 8.2% 2D FLAIR 1x1x2-5 |
| | | | | | T1w | | [1570-1660] / [2.54-3.42] / [800-850] | 49.4% 0.8x0.8x0.8, 50.6% 1x1x1 |
| | | Controls | - | 85 | T2w-FLAIR | 22 to 62 yrs, 33.3 ± 11.9 | 5000 / 388 / 1800 | 100% 3D FLAIR: 1x1x1; |
| | | | | | T1w | | [1570-1660] / [2.54-3.42] / [800-850] | 91.8% 0.8x0.8x0.8, 8.2% 1x1x1 |
| D2 | 3T General Electric (GE) DISCOVERY MR750, 3T GE SIGNA HDx | FCD Patients | Type IIa: 21.6%, Type IIb: 78.4% | 32 | T2w-FLAIR | 15 to 70 yrs, 28.3 ± 11.26 | - / - / [1874-2250] | [0.8 to 1.1] x [0.8 to 1.1] x [0.5 to 5.0] |
| | | | | | T1w | | - / - / [400-450] | [0.5 to 1.0] x [0.5 to 1.0] x [0.9 to 1.1] |
| D3 | 3T, Siemens, Different Models: Prisma, Verio, Skyra | FCD Patients | Type IIa: 57.1%, Type IIb: 42.9% | 14 | T2w-FLAIR | 1.5 to 20.8 yrs, 8.43 ± 6.45 | [5860-11750] / [89-135] / [2500] | [0.4 to 0.8] x [0.4 to 0.8] x [1.0 to 4.0] |
| | | | | | T1w | | [1950-2350] / [1.69-2.26] / [900-1450] | [0.5 to 1.0] x [0.5 to 1.0] x [0.8 to 1.1] |
| | | Controls | - | 5 | T2w-FLAIR | | [5860-11750] / [89-135] / [2500] | [0.4 to 0.8] x [0.4 to 0.8] x [1.0 to 4.0] |
| | | | | | T1w | | [1950-2350] / [1.69-2.26] / [900-1450] | [0.5 to 1.0] x [0.5 to 1.0] x [0.8 to 1.1] |

Three datasets are dataset-1 (D1), dataset-2 (D2), and dataset-3 (D3) have T1-weighted (T1w) and T2 weighted Fluid-Attenuated Inversion Recovery (FLAIR) images. Repetition time (TR) and echo time (TE) for Dataset-2 (D2) is not accessible. Here, TI is Inversion time. Age for the dataset is mentioned as mean ± std along with minimum (min) and maximum (max) values, measured in years (yrs).

**Table 2.** Performance evaluation of Focal Cortical Dysplasia detection for the Dataset-1 test cohort (15 patients and 15 controls) using reduced real data, synthetic data augmentation, and expanded real data across all three datasets.

| Level | Metric | FCD detector Trained on 35 patients + 35 controls | FCD detector Trained on 35 patients and 35 controls , and synthetic data | FCD detector Trained on 70 patients and 70 controls |
|---|---|---|---|---|
| Voxel | Dice Score | 0.3318 [0.31-0.35] | 0.2112 [0.19-0.24] | 0.4173 [0.39-0.42] |
| Cluster | Number of False Positive clusters / Number of subjects | 0.33 [0.18-0.48] | 0.78 [0.56-1.01] | 0.31 [0.46-0.88] |
| | Positive Predicted Value (True Positive Predicted/ Total Positive Predicted) | 0.59 (10/17) | 0.50 (11/22) | 0.57 (12/21) |
| | Sensitivity (True Positive Predicted/ Total Positive) | 0.67 (10/15) | 0.73 (11/15) | 0.8 (12/15) |
| | F1-score | 0.627 | 0.593 | 0.66 |
| | Dice score ( Voxelwise for Lesion region ) | 0.7393 | 0.4509 | 0.6875 |
| Subject | Pinpointing | 0.67 (10/15) | 0.73 (11/15) | 0.80 (12/15) |
| | Detection Rate | 0.53 (8/15) | 0.60 (9/15) | 0.80 (12/15) |
| | Negative Predicted Value (True Negative Predicted/ Total Negative Predicted) | 0.78 (14/18) | 0.73 (11/15) | 0.92 (12/13) |
| | Specificity | 0.93 (14/15) | 0.73 (11/15) | 0.80 (12/15) |

FCD: Focal Cortical Dysplasia

**Table 3.** Performance evaluation of Focal Cortical Dysplasia detection for the full test cohort ( 61 patients / 20 controls) using models trained on reduced real data, synthetic data augmentation, and expanded real data.

| Level | Metric | FCD detector Trained on 35 patients + 35 controls | FCD detector Trained on 35 patients and 35 controls , and synthetic data | FCD detector Trained on 70 patients and 70 controls |
|---|---|---|---|---|
| Cluster | Number of False Positive clusters / Number of subjects | 0.37 (0.22 - 0.52) | 0.57 (0.36 - 0.78) | 0.87 (0.63 - 1.13) |
| | Positive Predicted Value (True Positive Predicted/ Total Positive Predicted) | 0.49 (22/45) | 0.41 (26/63) | 0.44 (45/103) |
| | Sensitivity (True Positive Predicted/ Total Positive) | 0.36 (22/61) | 0.43 (26/61) | 0.74 (45/61) |
| Subject | Pinpointing | 0.36 (22/61) | 0.44 (27/61) | 0.74 (45/61) |
| | Detection Rate | 0.36 (22/61) | 0.41 (25/61) | 0.67 (41/61) |
| | Specificity | 0.95 (19/20) | 0.70 (14/20) | 0.75 (15/20) |

FCD: Focal Cortical Dysplasia

**Table 4.** Comparison of the proposed method with existing approaches for detecting Focal Cortical Dysplasia (FCD) including other structural brain anomalies, in terms of training data size, number of trainable parameters, test set size, accuracy, sensitivity, and specificity.

| Method Number | Methods | Pathology | Training Samples (Controls : Patients) | Trainable Parameters | Dataset Name/ Test Samples (Controls : Patients) | Accuracy | Sensitivity | Specificity |
|---|---|---|---|---|---|---|---|---|
| 1 | MELD-1 [7] | FCD | 180:278 | ~2k | Dataset-1 [16] / 15:15 | 0.40 | 0.60 | 0.20 |
| 2 | Proposed (reduced real+synthetic) | FCD | 35:35 + Augmentation | 17M | Dataset-1 [16] / 15:15 | 0.73 | 0.73 | 0.73 |
| 3 | Proposed (real expanded) | FCD | 70:70 | 17M | Dataset-1 [16] / 15:15 | 0.80 | 0.80 | 0.80 |
| 4 | MELD-1 [7] | FCD | 180:278 | ~2k | Dataset-3 / 14:5 | 0.58 | 0.57 | 0.60 |
| 5 | Proposed (reduced real+synthetic) | FCD | 35:35 | 17M | Dataset-3 / 14:5 | 0.58 | 0.50 | 0.80 |
| 6 | Proposed (real expanded) | FCD | 70:70 | 17M | Dataset-3 / 14:5 | 0.68 | 0.64 | 0.80 |

Multi-centre Epilepsy Lesion Detection (MELD) was additionally evaluated on Test Set-1 (Held-out from Dataset-1) for direct comparison with the proposed method.

CNN: Convolutional Neural Network. WMH: White Matter Hyperintensities. Trainable parameters: 10,000 is represented by 10k, and $17 \times 10^6$ is represented by 17M. FCD: Focal Cortical Dysplasia

**Figure 1:** Demonstration of generation of MRI volumes with Focal Cortical Dysplasia (FCD). Red circles in images indicate the region in which FCD was introduced, and are zoomed out for better visualization. Red arrows in the right most T1 weighted (T1w) and T2 weighted Fluid-Attenuated Inversion Recovery (FLAIR) images further highlight the location of FCD.

Box A illustrates the FCD label generator. Partial volume estimates are first computed, followed by extraction of white matter–gray matter (WM–GM) boundary voxels. A Gaussian mixture model is then applied to generate a binary mask indicating the presence of FCD.

Box B presents the image generator based on Spatially Adaptive Denormalization (SPADE). Partial volume maps of six brain tissues are used as inputs, together with a reference image providing the desired contrast. These inputs condition the SPADE-based generator, which synthesizes images by combining structural information from the partial volume maps with the contrast characteristics of the reference image.

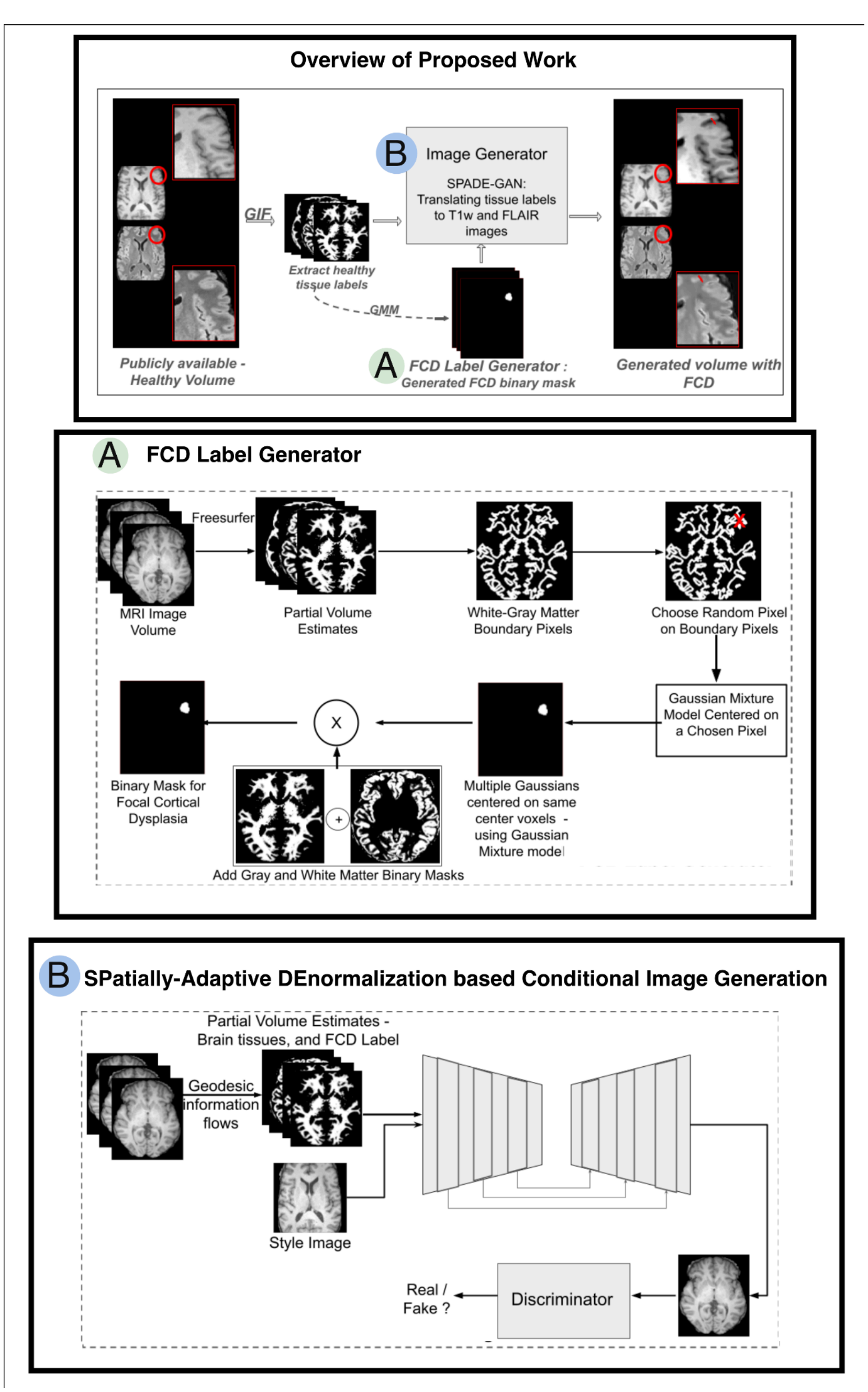

**Figure 2:** Graphical representation of experiment setup to assess the impact of synthetic data in automated Focal Cortical Dysplasia (FCD) detection in low data settings. Three models are shown, each trained with reduced real data (FCDD1), real data augmented with synthetic data (FCDD2), and expanded real data (FCDD3). FCDD1 and FCDD3 were compared to evaluate the impact of reducing the labeled training data. FCDD1 and FCDD2 were compared to evaluate the impact of synthetic data augmentation.

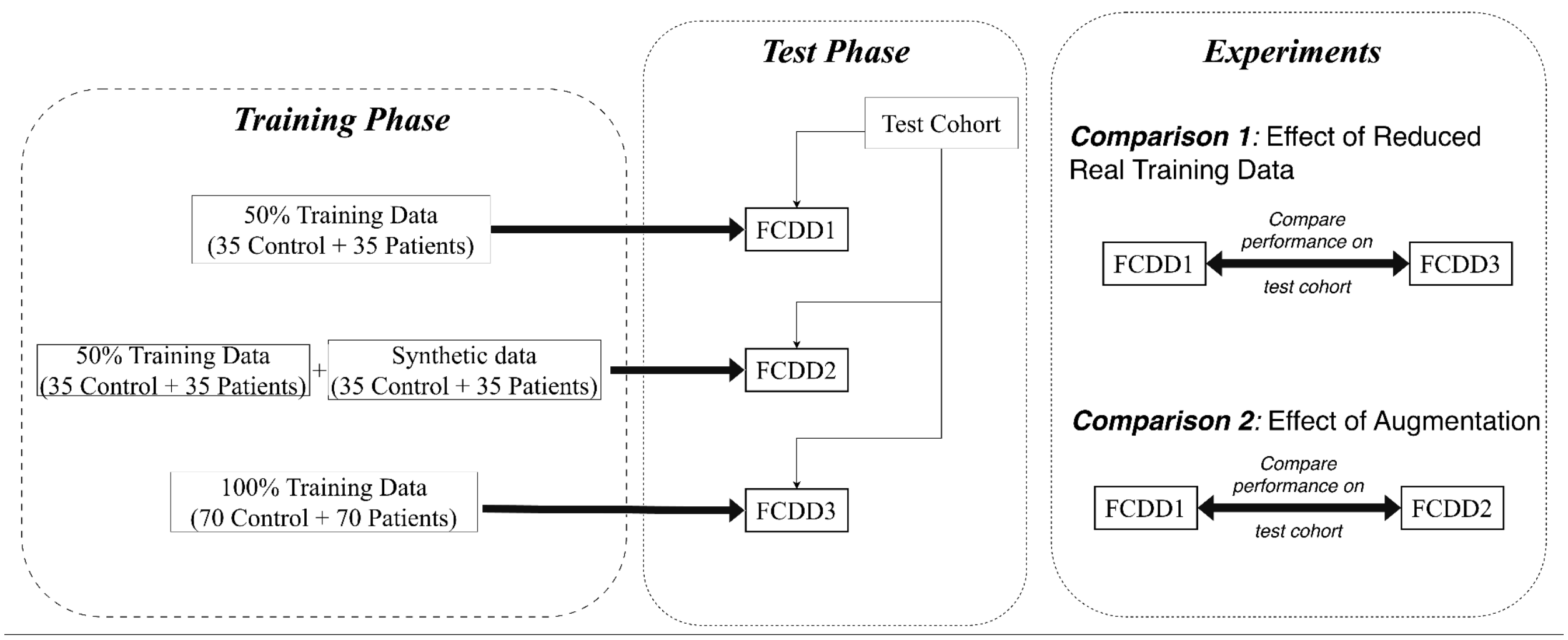

**Figure 3:** Example MRI volumes exhibiting focal cortical dysplasias (FCD) in real and synthetically generated scans. The red arrows indicate the FCD present in images. T1 weighted images, Fluid-Attenuated Inversion Recovery Images are represented by T1 and FLAIR.

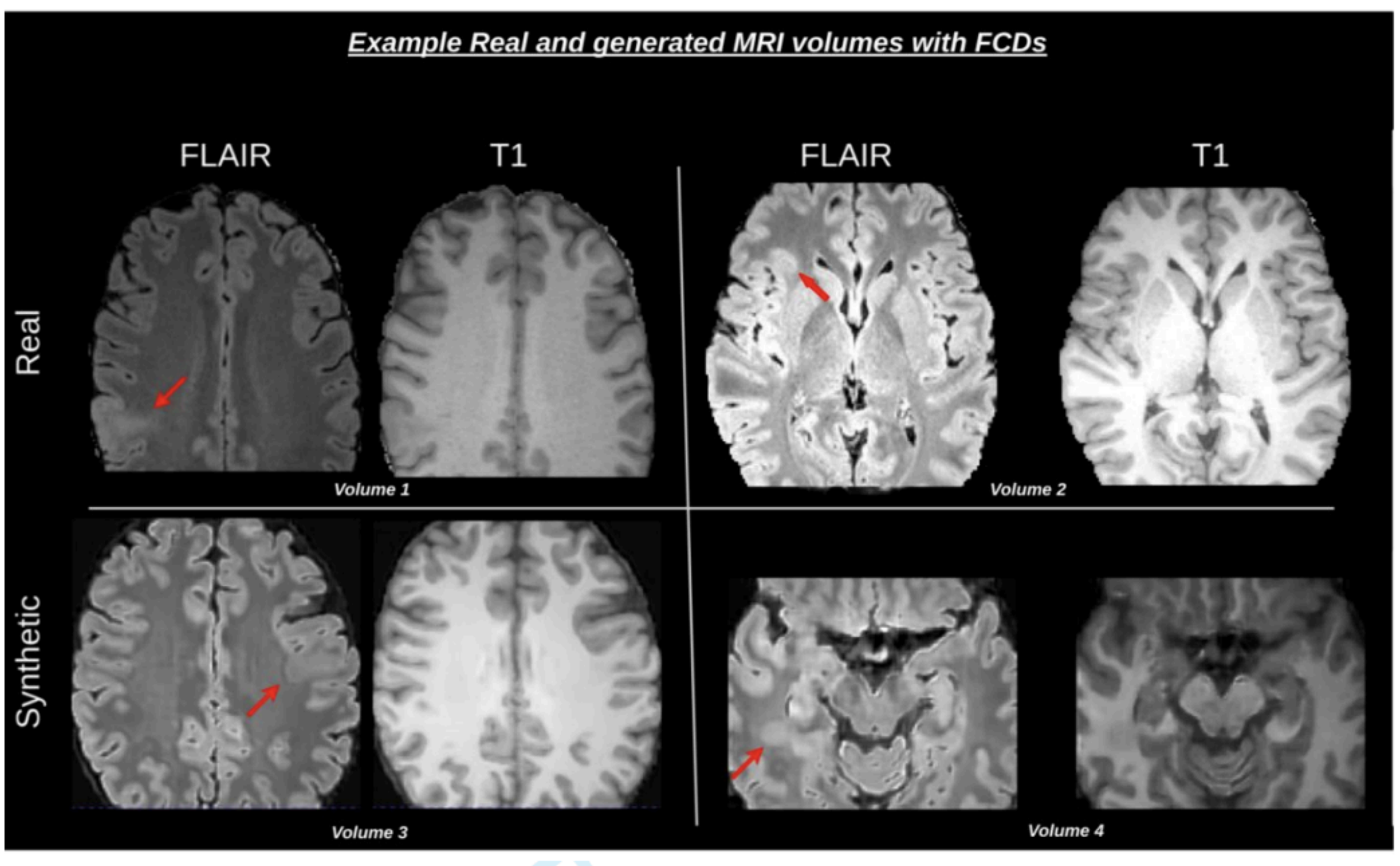

**Figure 4:** Illustration of examples where Focal cortical dysplasia (FCD) detector (FCDD) trained without synthetic data augmentation - FCDD1) correctly identified Focal Cortical Dysplasia cases missed by FCDD2 (trained with synthetic data augmentation), and vice versa. Two planes of each subject are displayed with and without segmentation to illustrate the imaging features of lesion.The outputs of FCDD2 for Row 2 volumes exhibited false-positives for normal T2 weighted-Fluid-Attenuated Inversion Recovery image hyperintensities. Lesion Size represents the number of voxels with lesion. Red blobs indicate the prediction, and are contoured with red boundaries in zoomed versions for better visualization.

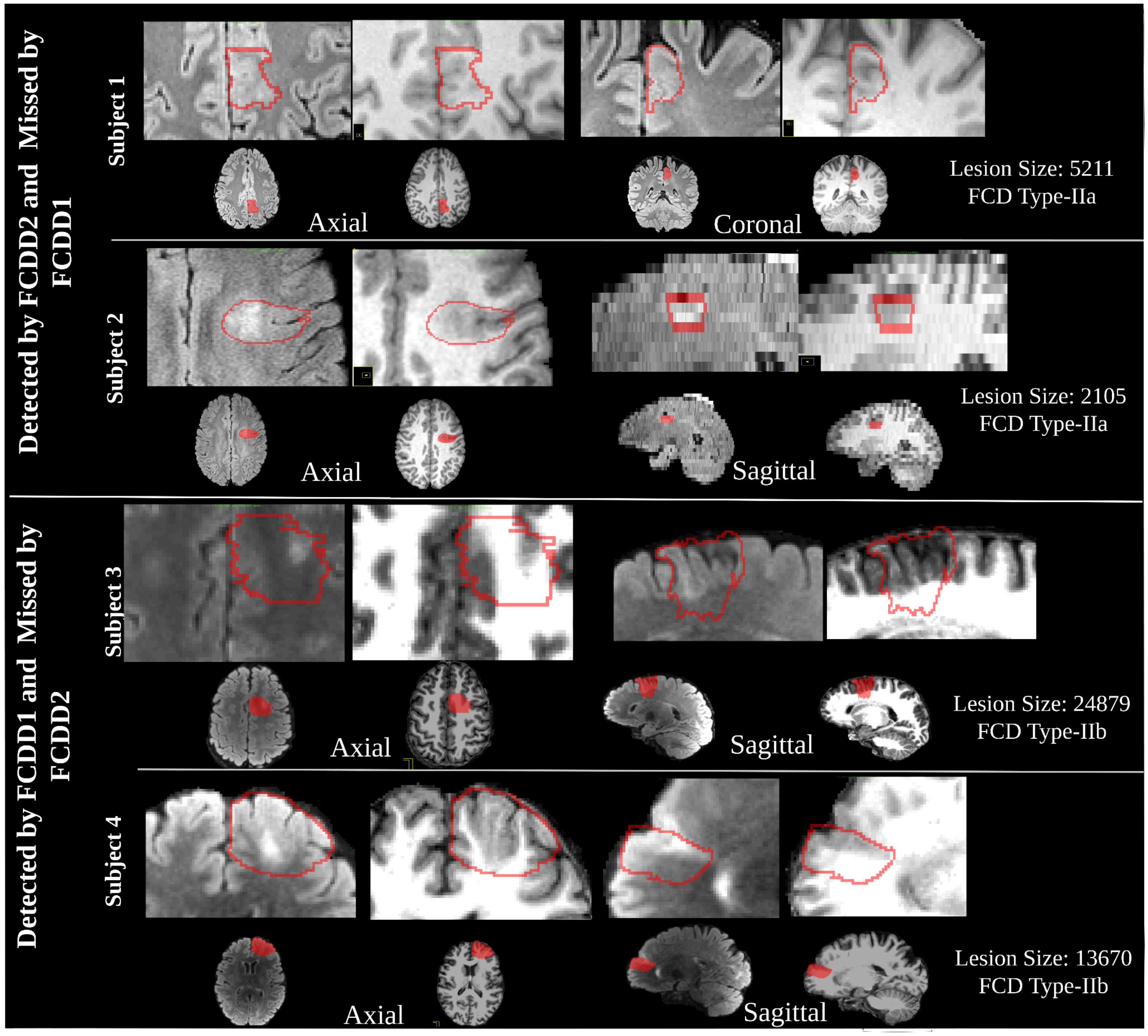

**Figure 5.** Qualitative Comparison of Multi-centre Epilepsy Lesion Detection (MELD) [7] with neural network based on encoder-decoder architecture with skip connections (nnU-Net) method to detect Focal Cortical Dysplasia (FCD).

MELD is trained on the largest curated FCD dataset (278 patients and 180 controls) as compared to nnU-Net trained on 70 FCD patients+70 controls. Subject 1 is a patient with FCD and Subject 2 is a healthy subject. This figure demonstrates the prediction of two methods for a patient and a healthy subject. The first row shows true positive predictions by the methods, while the second row shows false positives. The nnU-Net method showed less false positives and more true positives as compared to MELD, this indicates that selecting nnU-Net method to evaluate synthetic data is an optimal choice.

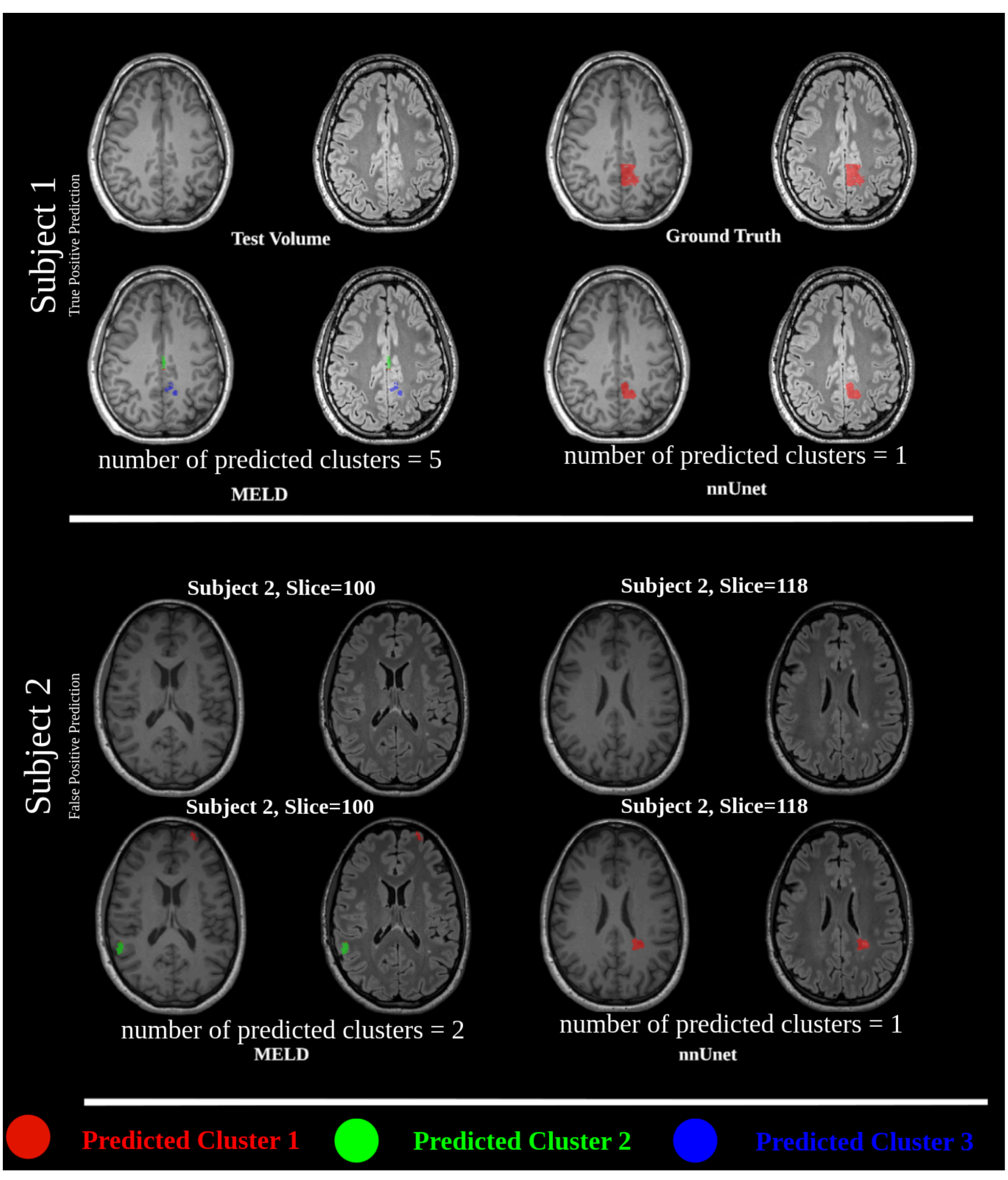

**Figure 6.** Diversity of synthetic Focal Cortical Dysplasia (FCD) MRI images in terms of location and FCD size (count representing number of voxels).

A. Regional distribution of FCD lesions across the brain lobes

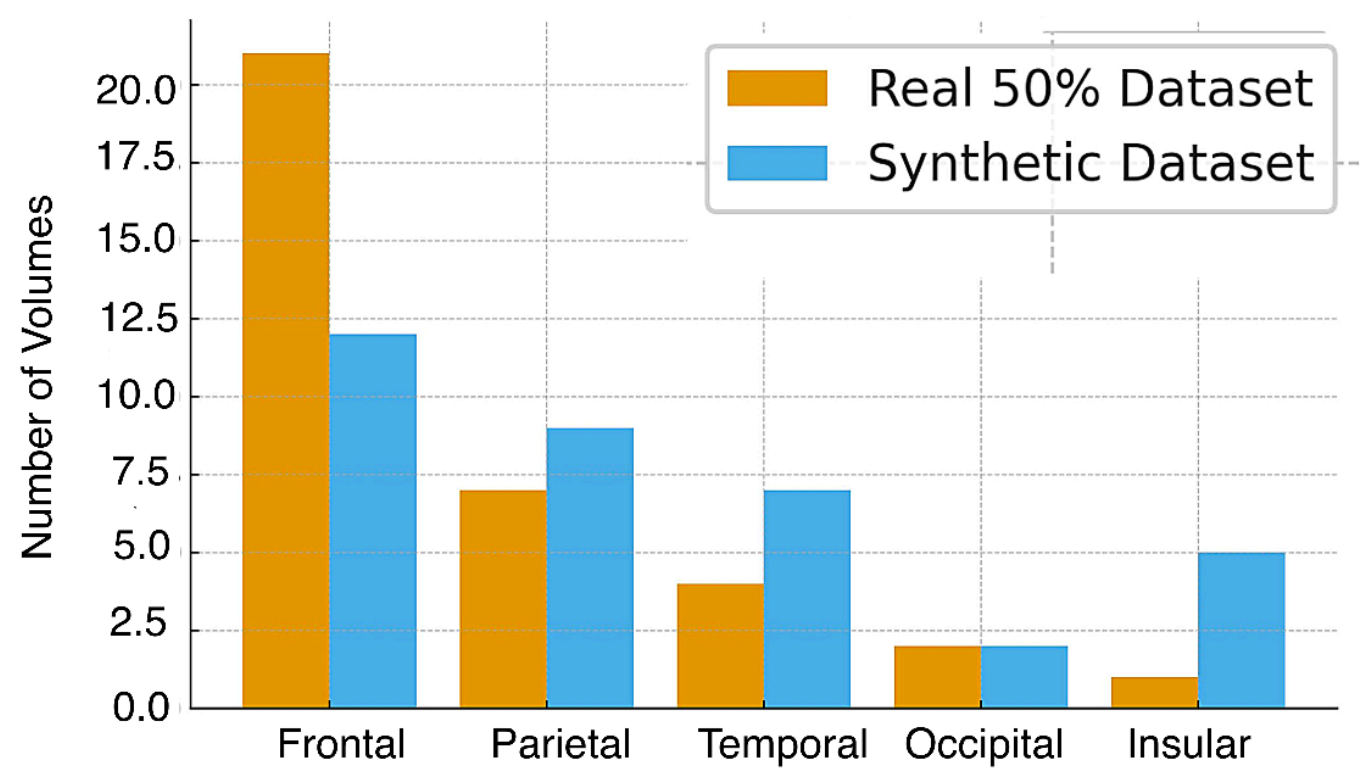

B. Distribution of lesion size comparison between real data and synthetic data

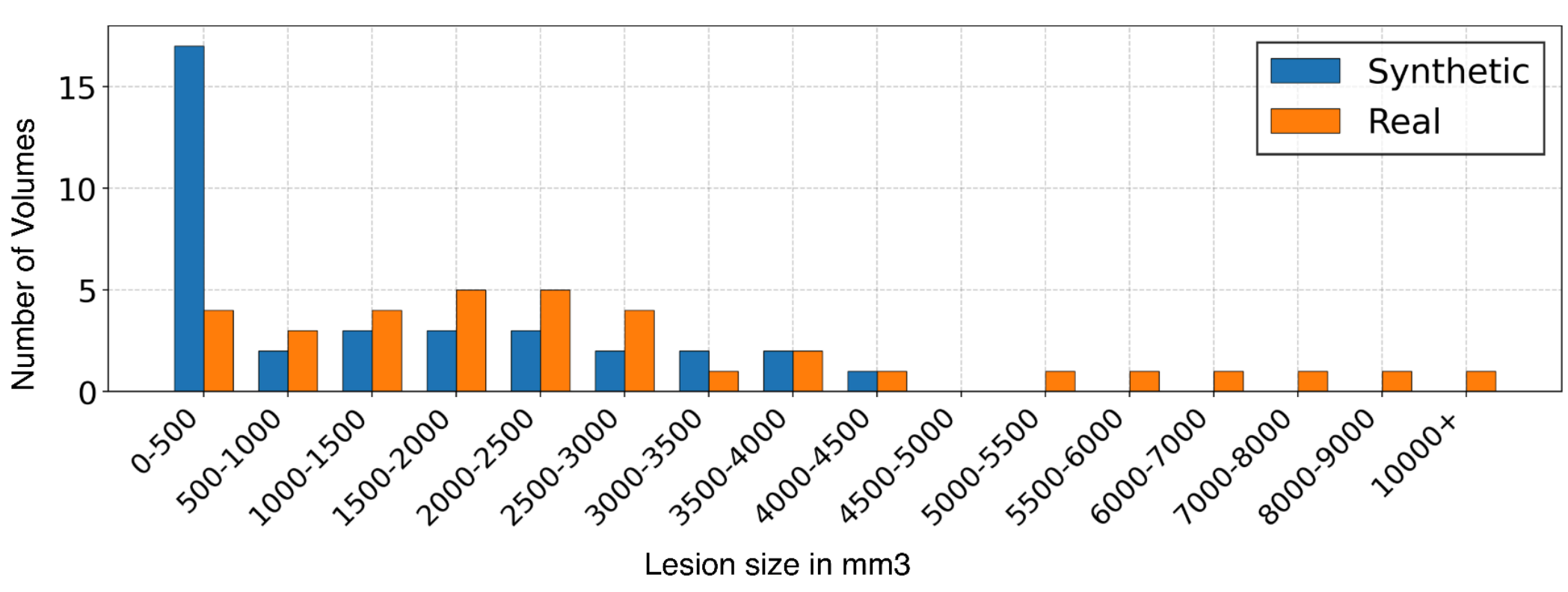